\documentclass[11pt]{article}
\usepackage{amsmath,amssymb,amsthm}
\usepackage{mathtools}
\usepackage{graphics,epsfig}
\usepackage{hyperref}
\usepackage{natbib}
\usepackage{color}
\usepackage{graphicx}
\usepackage{caption}
\usepackage{subcaption}
\usepackage{float}
\usepackage{mathrsfs}
\usepackage{multirow}
\usepackage{comment}
\usepackage{setspace}
\usepackage{bbm}
\usepackage{bm}
\usepackage{amsfonts}
\usepackage{xcolor}
\usepackage{pslatex}
\usepackage{setspace}
\usepackage{bbm}
\usepackage{booktabs}
\usepackage{listings}


\begin{document}

\title{\bfseries Obesity and Sociodemographic Factors in Luminal Breast Cancer}

\author{
 Vacanti Anderson$^{1}$ \and 
    Paramahansa Pramanik$^{2,4}$ \and 
    Haley K. Robinson$^{3}$
}

\date{
    \small
    $^{1}$ School of Computing, University of South Alabama, Mobile, AL 36688, United States.\\
    \texttt{vaa2021@jagmail.southalabama.edu}\\
    \vspace{0.5em}
    $^{2}$ Department of Mathematics and Statistics, University of South Alabama, Mobile, AL 36688, United States.\\
\vspace{0.5em}
    $^{3}$ Department of Biomedical Sciences,
University of South Alabama,
Mobile, Alabama 36688, United States.
\texttt{hkr2322@jagmail.southalabama.edu}\\
    $^{4}$Corresponding author, \texttt{ppramanik@southalabama.edu}
}

%\date{\today}
\maketitle

\begin{abstract}
Luminal breast cancers represent the most prevalent molecular subtype of breast carcinoma, with Luminal A tumors generally associated with more favorable clinical outcomes than Luminal B tumors. Obesity-related inflammation and prolonged exposure to exogenous steroids have been implicated in the progression of luminal malignancies. This study evaluated 1,928 patients with Luminal A breast cancer and 1,610 patients with Luminal B breast cancer to examine associations among body mass index (BMI), age, ethnic background, menopausal status, and receptor expression, including estrogen receptor (ER), progesterone receptor (PR), and human epidermal growth factor receptor 2 (HER2). Patients with Luminal B tumors demonstrated a significantly greater mean BMI compared with those with Luminal A tumors \((\Delta = 0.69\ \text{kg/m}^2,\ 95\%\ \text{CI}: 0.17\text{--}1.21,\ p = 0.010)\). In addition, Luminal B tumors were more frequently observed among patients of African ancestry relative to White and Hispanic populations \((p \leq 0.001)\). Multivariable analyses revealed that elevated BMI \((p = 0.008)\) and African ancestry \((p < 0.001)\) were independently associated with increased odds of Luminal B carcinoma, whereas postmenopausal status was associated with lower risk \((p = 0.028)\). Mediation analysis further indicated that BMI partially explained the association between ancestry and Luminal B disease. These findings suggest that obesity and population-specific factors may contribute to the development of more aggressive luminal breast cancer phenotypes.
\end{abstract}

{\bf Keywords:} Breast cancer; obesity; ethnicity; Luminal A; Luminal B.

\section{Introduction:}
This study investigates the relationship between obesity, ethnic background, and luminal subtypes of breast cancer to improve understanding of variations in disease presentation and outcomes \citep{hertweck2023clinicopathological,khan2023myb}. Luminal A and Luminal B tumors differ in their molecular characteristics, clinical behavior, and response to therapy, and these differences may be influenced by both genetic and environmental factors. Previous studies have shown that obesity is associated with a higher risk and poorer survival in hormone receptor-positive breast cancers \citep{kakkat2023cardiovascular,khan2023myb}. In addition, patient demographic factors have been linked to differences in breast cancer diagnosis, treatment patterns, and outcomes, independent of age, socioeconomic conditions, and coexisting illnesses. Evaluating the association between obesity, population background, and luminal breast cancer subtypes may provide important insight into disease disparities and support the development of more personalized prevention strategies and treatment approaches \citep{kakkat2023cardiovascular,khan2024mp60}.

Breast cancer continues to be one of the primary causes of cancer-related death among women globally. Epidemiologic evidence has consistently identified obesity as an important contributor to breast cancer risk, particularly among postmenopausal women. Despite advances in screening and treatment, substantial differences in incidence patterns and clinical outcomes persist across population groups. Variations in tumor biology, disease progression, and survival have been observed among patients from different ethnic and demographic backgrounds, especially within luminal breast cancer subtypes such as Luminal A and Luminal B. This study explores the relationship between obesity and population background in the development of luminal breast cancer subtypes and evaluates how these factors may collectively influence disease characteristics and prognosis \citep{maki2025new}. Luminal breast cancers are commonly characterized by estrogen receptor (ER) positivity, comprise approximately 70\% of all breast cancer diagnoses. Luminal A tumors are typically positive for both ER and progesterone receptor (PR) expression while lacking HER2 overexpression, whereas Luminal B tumors generally demonstrate ER positivity with HER2 expression and reduced or absent PR expression \citep{vikramdeo2024abstract,vikramdeo2023profiling}. These subtypes are distinguished by differences in two major biologic mechanisms: proliferation-associated signaling pathways and luminal regulatory pathways. Variability in these molecular processes contributes to the distinct clinical behavior, treatment response, and prognosis observed between Luminal A and Luminal B breast cancers.

Breast cancers associated with obesity, defined as a body mass index (BMI) of at least 30\,kg/m\textsuperscript{2}, are more frequently linked to estrogen receptor (ER) positive and progesterone receptor (PR) positive tumors than to ER-negative disease, particularly among postmenopausal women resulting, these tumors are often more responsive to endocrine-based therapies \citep{pramanik2024bayes}. Obesity is a complex chronic condition characterized by multiple physiologic and molecular disturbances, including impaired metabolic regulation, insulin resistance, oxidative stress, endoplasmic reticulum stress, and increased production of pro-inflammatory cytokines. Together, these alterations contribute to a biologic environment that may promote tumor initiation and progression \citep{khan2024mp60,dasgupta2023frequent,hertweck2023clinicopathological}. Understanding the influence of these obesity-related metabolic changes on proliferation-associated and luminal-regulated signaling pathways is essential within the framework of molecular breast cancer profiling. Furthermore, the contribution of obesity-driven inflammation and population-specific factors should be evaluated in the genomic context of breast cancer development \citep{pramanik2021optimala,pramanik2021scoring}. In the present study, we investigated the relationship between obesity and genomically characterized breast cancer subtypes. Our findings provide insight into molecular pathways potentially altered by obesity-associated pathophysiologic mechanisms and may help clarify disparities observed among different population groups while supporting more individualized strategies for breast cancer prevention and treatment \citep{kakkat2026angiotensin}.

Our study was motivated by increasing evidence that obesity and population background may jointly influence the development and progression of breast cancer, particularly among luminal molecular subtypes. Although obesity is widely recognized as a major risk factor for ER positive and PR positive breast cancers, its effects across different demographic populations have not been fully characterized \citep{pramanik2020optimization,pramanik2023semicooperation}. Previous investigations have demonstrated that obesity disproportionately affects certain communities and is associated with differences in breast cancer incidence, tumor characteristics, and clinical outcomes among diverse populations. In addition, the biologic mechanisms linking obesity to breast cancer progression, especially within Luminal A and Luminal B tumors, remain incompletely understood. Because these luminal subtypes differ substantially in molecular features, therapeutic response, and prognosis, evaluating the influence of obesity across varying population groups may improve understanding of disparities in breast cancer outcomes \citep{pramanik2020motivation}. Accordingly, this study examines the associations among obesity, population background, and the molecular characteristics of luminal breast cancer subtypes with the goal of supporting more individualized prevention strategies and therapeutic approaches.

\subsection{Background Literature:}
 The relationship between obesity, population background, and luminal breast cancer subtypes has become an increasingly important area of investigation because of its potential role in explaining disparities in breast cancer incidence, progression, and survival outcomes across diverse patient populations. Obesity is widely recognized as a major contributor to the development of several malignancies, including breast cancer, and substantial epidemiologic evidence has demonstrated its involvement in tumor initiation, disease progression, and overall prognosis \citep{Calle2003, Renehan2008}. However, the influence of obesity on breast cancer risk is not consistent across all demographic groups, suggesting that biologic susceptibility may interact with environmental, socioeconomic, and healthcare-related factors \citep{Banegas2014, Hastert2016}. These observations indicate that the relationship between obesity and breast cancer extends beyond simple metabolic effects and may reflect a broader interaction between physiologic processes and social determinants of health. As a result, understanding how obesity contributes to differences in breast cancer outcomes among varying populations remains essential for improving prevention strategies, early detection, and individualized patient management.

Breast cancer is classified into several intrinsic molecular subtypes according to hormone receptor expression and HER2 status, with Luminal A and Luminal B representing the most common hormone receptor--positive categories \citep{Prat2015}. Although both subtypes are generally characterized by estrogen receptor positivity, they differ substantially in biologic behavior, proliferation activity, treatment response, and long-term clinical outcomes. Luminal A tumors are typically associated with slower growth patterns, lower recurrence rates, and improved responsiveness to endocrine therapy, resulting in more favorable prognoses. In contrast, Luminal B tumors commonly demonstrate increased proliferative activity, greater biologic aggressiveness, higher recurrence risk, and poorer survival outcomes, often requiring more intensive therapeutic approaches that may include chemotherapy in addition to hormonal treatment \citep{Carey2006, Sorlie2001}. These molecular and clinical distinctions highlight the importance of accurately identifying factors that may contribute to the development of one luminal subtype over another, particularly in populations that may already experience disparities in breast cancer care and outcomes.

Accumulating evidence suggests that obesity may differentially influence the development and progression of luminal breast cancer subtypes through several interconnected biologic mechanisms. Adipose tissue functions not only as an energy storage organ but also as an active endocrine organ capable of producing estrogens through aromatization and releasing inflammatory mediators that can promote tumor growth and progression \citep{Iyengar2016, Park2017}. Increased body mass index (BMI) has been associated with a greater likelihood of developing Luminal B breast cancer, especially among postmenopausal women \citep{Pierobon2013, Phipps2011}. In addition, obesity-related chronic inflammation, insulin resistance, and altered metabolic signaling may contribute to the more aggressive clinical features frequently observed in Luminal B tumors \citep{Rose2015}. These obesity-associated physiologic disturbances may create a tumor-supportive microenvironment that enhances cellular proliferation and disease progression. Consequently, evaluating the relationship between obesity and luminal breast cancer subtypes may provide valuable insight into the biologic pathways underlying disparities in breast cancer behavior and may support the development of more targeted prevention and treatment strategies across diverse patient populations.

Differences across population groups add further complexity to the relationship between obesity and luminal breast cancer subtypes. Previous investigations have reported that women of African ancestry are more frequently diagnosed with Luminal B breast tumors when compared with non-Hispanic White and Hispanic women \citep{Carey2006, Amirikia2011}. This observation is clinically important because Luminal B tumors are generally associated with greater proliferative activity, more aggressive disease behavior, and less favorable clinical outcomes than Luminal A tumors. Multiple studies have suggested that these disparities may be partially explained by differences in obesity prevalence, since obesity rates tend to be higher among certain demographic populations and are associated with metabolic and inflammatory changes that may promote tumor progression \citep{Joslyn2002, Palmer2014}. In addition to differences in body composition and metabolic health, variations in tumor biology, patterns of healthcare utilization, and access to timely screening and treatment services may also contribute to the unequal distribution of breast cancer subtypes observed across populations. These findings emphasize the importance of examining both biologic and nonbiologic contributors when evaluating disparities in breast cancer incidence and prognosis.

The influence of obesity on breast cancer outcomes cannot be fully understood without considering the broader social and environmental context in which patients live. Socioeconomic conditions, neighborhood resources, environmental exposures, nutritional access, healthcare availability, and long-standing structural inequalities may all interact to influence both obesity prevalence and breast cancer outcomes across different communities \citep{DeSantis2019}. Limited access to preventive healthcare services and delayed diagnosis may contribute to more advanced disease presentation in certain populations, while chronic exposure to environmental and psychosocial stressors may further affect physiologic pathways associated with inflammation and cancer progression. In addition, disparities in treatment access, quality of care, and participation in clinical research may also influence survival outcomes among patients diagnosed with luminal breast cancer subtypes \citep{pramanik2024estimation,vikramdeo2024mitochondrial}. These overlapping factors illustrate that breast cancer disparities are multifactorial and likely result from a combination of biologic susceptibility, metabolic health, healthcare inequities, and environmental influences rather than a single isolated cause.

A more comprehensive understanding of the relationship among obesity, population background, and luminal breast cancer subtypes is essential for improving prevention strategies, screening approaches, and therapeutic management. Public health interventions focused on reducing obesity may provide substantial benefits, particularly among patient populations at increased risk for aggressive breast cancer phenotypes \citep{Tseng2018}. Furthermore, continued investigation into the molecular mechanisms linking obesity-related metabolic alterations with Luminal A and Luminal B tumors may help identify novel therapeutic pathways and improve precision-based treatment strategies \citep{pramanik2024motivation}. Future studies should continue to evaluate the combined influence of biologic, environmental, clinical, and social determinants on breast cancer development and progression. Expanding research efforts in diverse patient populations may improve understanding of disease heterogeneity and support the development of more equitable and individualized approaches to breast cancer prevention, diagnosis, and treatment \citep{bulls2025assessing}.

\subsection{Our Contribution:}
Our study examining the relationship among obesity, population background, and luminal breast cancer subtypes contributes important clinical and methodological insight to the current literature. A major strength of this research is the inclusion of a more demographically diverse patient population compared with many earlier studies that primarily focused on White patient cohorts, thereby allowing a broader evaluation of how obesity and population-specific factors may influence breast cancer subtype distribution and outcomes across different communities \citep{pramanik2024estimation,pramanik2023cont}. In addition, our analyses accounted for several potential confounding variables, including socioeconomic conditions and lifestyle-related factors, which have not always been comprehensively addressed in prior investigations. By specifically focusing on Luminal A and Luminal B breast cancers, this study further expands understanding of the differential impact of obesity on distinct hormone receptor--positive tumor subtypes that remain relatively underexplored in existing research. Our findings also highlight the importance of future longitudinal investigations to better characterize the long-term influence of obesity on breast cancer progression, particularly among historically underrepresented populations, as well as the need for additional genomic and molecular studies examining why certain demographic groups may be more susceptible to aggressive luminal tumor phenotypes in the setting of obesity \citep{pramanik2024estimation1,yusuf2025prognostic}. From a public health perspective, this work advances understanding of the interaction between obesity and population-related factors in breast cancer risk and may contribute to the development of more individualized prevention strategies and treatment approaches. Methodologically, the study incorporated extensive data preprocessing and quality assessment using R to ensure consistency and reliability across datasets. Statistical analyses included logistic regression modeling, survival analysis, and stratified analyses to evaluate associations among obesity, demographic characteristics, and luminal breast cancer subtypes. In addition, visualization tools implemented through R packages such as \texttt{ggplot2} and \texttt{survival} were used to generate Kaplan--Meier survival curves and forest plots that effectively summarized and illustrated the primary study findings.

\section{Methods:}
Some descriptive statistics were implemented to characterize the distribution of obesity, demographic background, and luminal breast cancer subtypes within the study population. Comparisons of categorical variables were performed using $\chi^2$ or Fisher’s exact F-tests when appropriate, whereas continuous variables were evaluated using independent sample t-tests or Wilcoxon rank-sum tests based on data distribution characteristics \citep{pramanik2023path}. To investigate the relationship between obesity and luminal breast cancer subtypes across different population groups, logistic regression was conducted with adjustment for potential confounding variables, including age, tumor stage, socioeconomic indicators, and coexisting medical conditions. Overall survival and progression-free survival were estimated using Kaplan-Meier survival analysis, and differences between groups were assessed with log-rank testing \citep{pramanik2023cmbp,yusuf2025predictive}. Cox proportional hazards regression models were subsequently used to estimate hazard ratios (HRs) and corresponding 95\% confidence intervals (CIs) for mortality outcomes while controlling for relevant clinical and demographic covariates. Potential interaction effects related to demographic background were additionally evaluated through interaction terms incorporated into the regression models. All statistical analyses were performed using R software (version 4.4.3), and statistical significance was defined as \(p < 0.05\).

This analysis compared patients with Luminal A and Luminal B breast cancer according to several clinical and demographic variables, including BMI, age, menopausal status, and population background. Group differences were evaluated using Welch’s t-tests with Welch-Satterthwaite correction for continuous variables and $\chi^2$tests with Yates’ continuity correction for categorical variables \citep{pramanik2021consensus}. To further examine factors associated with Luminal B breast cancer, univariate logistic regression analyses were performed using each variable as an independent predictor. Based on previous evidence suggesting demographic differences in breast cancer subtype prevalence, we also assessed whether the proportion of patients of African ancestry differed significantly between the Luminal A and Luminal B groups. Subsequently, all study variables were incorporated into a multivariable logistic regression model. Model selection was performed using a forward stepwise approach implemented through the \texttt{caret} package in R, and the final model was selected according to the lowest Akaike Information Criterion (AIC), indicating the best overall model fit \citep{pramanik2021,pramanik2022stochastic}.

We hypothesized that BMI might mediate the association between African ancestry and the likelihood of developing Luminal B breast cancer. To evaluate this possibility, mediation analysis was performed using Sobel’s method, and both direct and indirect effects through BMI were estimated. Confidence intervals for these effects were calculated using a nonparametric bootstrap procedure with 1,000 iterations. An important assumption underlying causal mediation analysis is sequential ignorability, which assumes the absence of unmeasured confounding variables affecting both the mediator (BMI) and the outcome (Luminal B breast cancer). Because this assumption cannot be directly verified using retrospective archival data, sensitivity analyses were conducted to assess the stability and robustness of the mediation findings \citep{pramanik2023optimization001}. These analyses incorporated a hypothetical unmeasured confounder with varying correlations between BMI and Luminal B breast cancer using sensitivity parameter $\rho\in[-1,1]$ \citep{imai2010identification,pramanik2025construction,pramanik2025optimal}. In addition, an exploratory mediation analysis was performed to determine whether the association between African ancestry and Luminal B breast cancer persisted after adjustment for menopausal status. Exploratory univariate and multivariable logistic regression analyses were also conducted to examine the relationship between estrogen receptor, progesterone receptor, and HER2-neu receptor positivity and the likelihood of Luminal B breast cancer relative to Luminal A disease.

\section{Data Analysis:}

\subsection{Source:}
The FLEX registry (NCT03053193) is a large, multicenter observational study developed to improve understanding of breast cancer through the integration of genomic profiling with real-world clinical information collected from routine patient care. Between December 14, 2018, and September 5, 2020, a total of 4,530 patients were enrolled from 90 participating institutions across the United States, creating a broad and diverse study population representative of contemporary breast cancer practice. Prior to patient enrollment, the study protocol underwent extensive review and approval by institutional review boards at all participating centers to ensure adherence to established ethical and scientific standards \citep{pramanik2024stochastic,pramanik2025stubbornness}. The investigation was conducted in accordance with the principles outlined in the Declaration of Helsinki, which provides internationally recognized guidance for the protection of research participants and the ethical conduct of human subject research. All participants provided written informed consent before inclusion in the registry, permitting the collection of clinical information, molecular profiling data, and the future dissemination of research findings derived from the study. Eligibility criteria required patients to have a historically confirmed diagnosis of early-stage breast cancer classified as stage I through stage III \citep{pramanik2025dissecting,pramanik2025factors}. In addition, all enrolled patients underwent MammaPrint and BluePrint genomic assays as part of their routine clinical management. These molecular profiling platforms provide detailed information regarding tumor biology and genomic characteristics, thereby supporting individualized treatment planning and improving prognostic assessment.

The registry included patients ranging in age from 18 to 90 years, allowing representation across a wide spectrum of demographic and clinical backgrounds. Beyond standard clinical information, participants also consented to the collection of clinically annotated full-transcriptome tumor profiling data, enabling a more comprehensive investigation of the molecular features associated with breast cancer development and progression \citep{pramanik2021thesis,pramanik2016}. This extensive genomic characterization provides researchers with an opportunity to better understand the biologic diversity of breast cancer and to explore molecular pathways that may influence treatment response and long-term outcomes. Importantly, all therapeutic decisions remained under the direction of the treating physicians, ensuring that patient care was individualized according to current clinical guidelines, physician judgment, and patient-specific considerations rather than dictated by the study protocol. Consequently, the FLEX registry represents an important real-world resource for advancing precision oncology by combining genomic testing with routine clinical practice data \citep{hua2019assessing,polansky2021motif}. The registry therefore offers valuable evidence that may contribute to the refinement of personalized treatment strategies, improvement of prognostic evaluation, and optimization of breast cancer management across diverse patient populations.

\subsection{Molecular and Clinical Subtype Analysis:}
MammaPrint and BluePrint are genomic profiling assays designed to classify breast cancer according to underlying molecular characteristics using microarray-based gene expression analysis. Both tests are performed exclusively at the Agendia Laboratory located in Irvine, California, USA, and are widely used to provide clinically relevant prognostic and predictive information that may assist physicians in treatment planning and therapeutic decision-making \citep{pramanik2024estimation,pramanik2023cont}. MammaPrint is based on a 70-gene expression signature developed to evaluate the likelihood of distant recurrence in patients with breast cancer. Through analysis of tumor gene expression patterns, the assay categorizes tumors into distinct risk groups that may help guide the intensity of treatment approaches. Tumors with a \(\text{MammaPrint index} > 0\) are classified as low risk, indicating a lower probability of metastatic recurrence and the potential to avoid unnecessary chemotherapy exposure \citep{pramanik2025strategic,pramanik2025impact}. In contrast, tumors with a \(\text{MammaPrint index} \geq 0\) are categorized as high risk, reflecting an increased likelihood of recurrence and a possible need for more intensive therapeutic management.

These genomic assays provide important insight into tumor biology beyond traditional clinicopathologic assessment and support a more individualized approach to breast cancer care. By evaluating gene expression patterns associated with tumor aggressiveness and recurrence risk, MammaPrint and BluePrint contribute to improved risk stratification and may help clinicians determine whether patients are likely to benefit from chemotherapy in addition to endocrine or targeted therapies \citep{pramanik2025strategies,pramanik2023optimization001}. The molecular information generated from these assays may also reduce overtreatment in patients with lower-risk disease while identifying individuals who may require closer monitoring or more aggressive intervention because of higher-risk tumor characteristics. As precision oncology continues to evolve, genomic profiling tools such as MammaPrint and BluePrint have become increasingly valuable in supporting evidence-based clinical decisions and optimizing treatment selection according to the biologic behavior of individual breast tumors.

\begin{figure}[htbp]
    \centering
\includegraphics[width=0.9\textwidth]{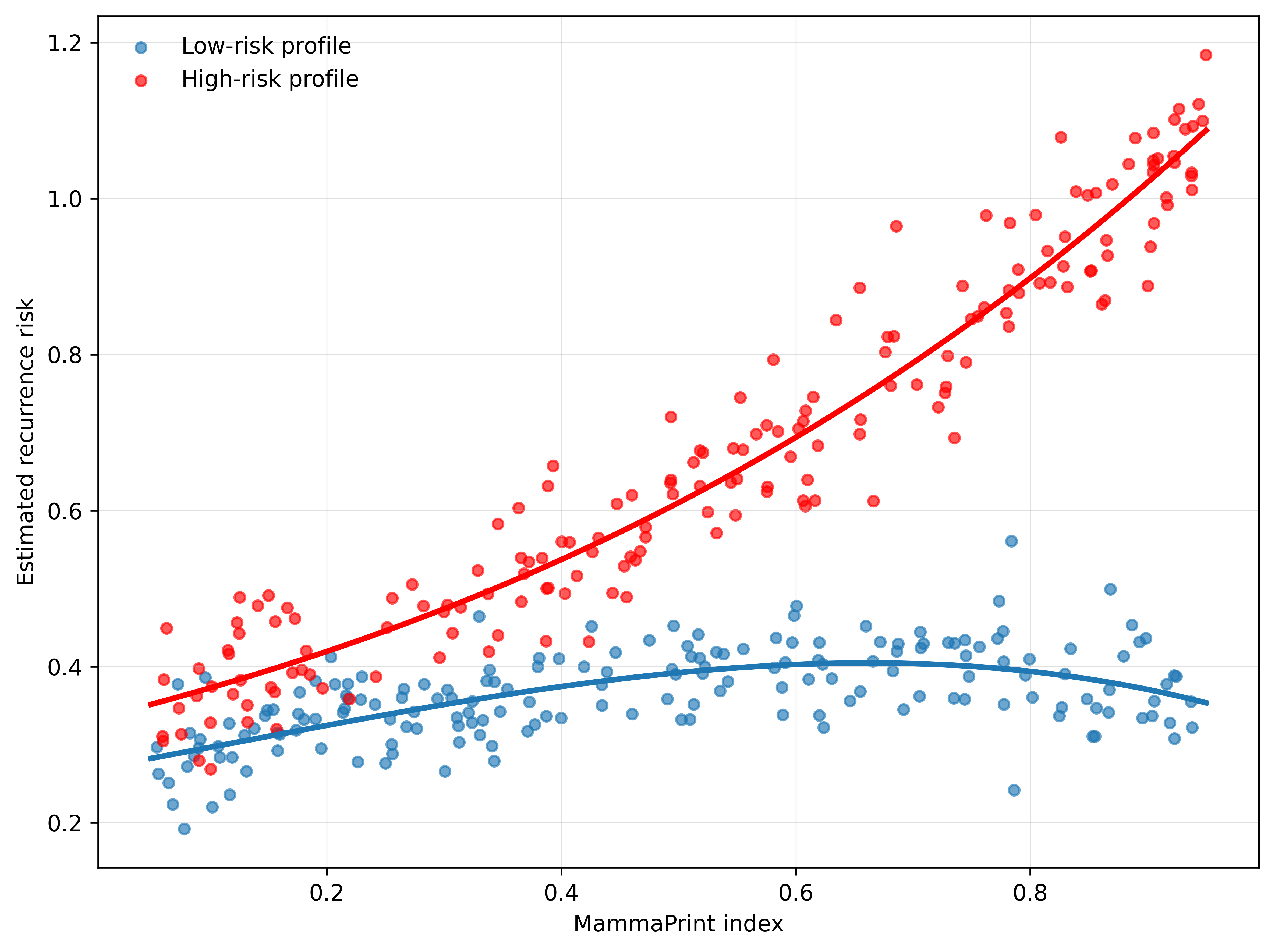}
    \caption{Nonlinear fitted trends between MammaPrint index and estimated recurrence risk profiles.}
    \label{fig:scatterplot}
\end{figure}
In Figure \ref{fig:scatterplot}, the scatterplot illustrates the relationship between the MammaPrint index and estimated recurrence risk across two molecular risk profiles. The low-risk profile demonstrates a relatively stable nonlinear trend with only modest variation in recurrence risk as the MammaPrint index increases \citep{pramanik2024measuring,pramanik2024dependence}. In contrast, the high-risk profile exhibits a pronounced upward nonlinear pattern, indicating that higher MammaPrint index values are associated with substantially greater estimated recurrence risk. The fitted nonlinear trend curves further emphasize the distinct progression patterns between the two groups. Overall, the figure highlights the differing biological behavior and recurrence tendencies associated with low-risk and high-risk molecular breast cancer profiles.

\begin{figure}[htbp]
    \centering
    \includegraphics[width=\textwidth]{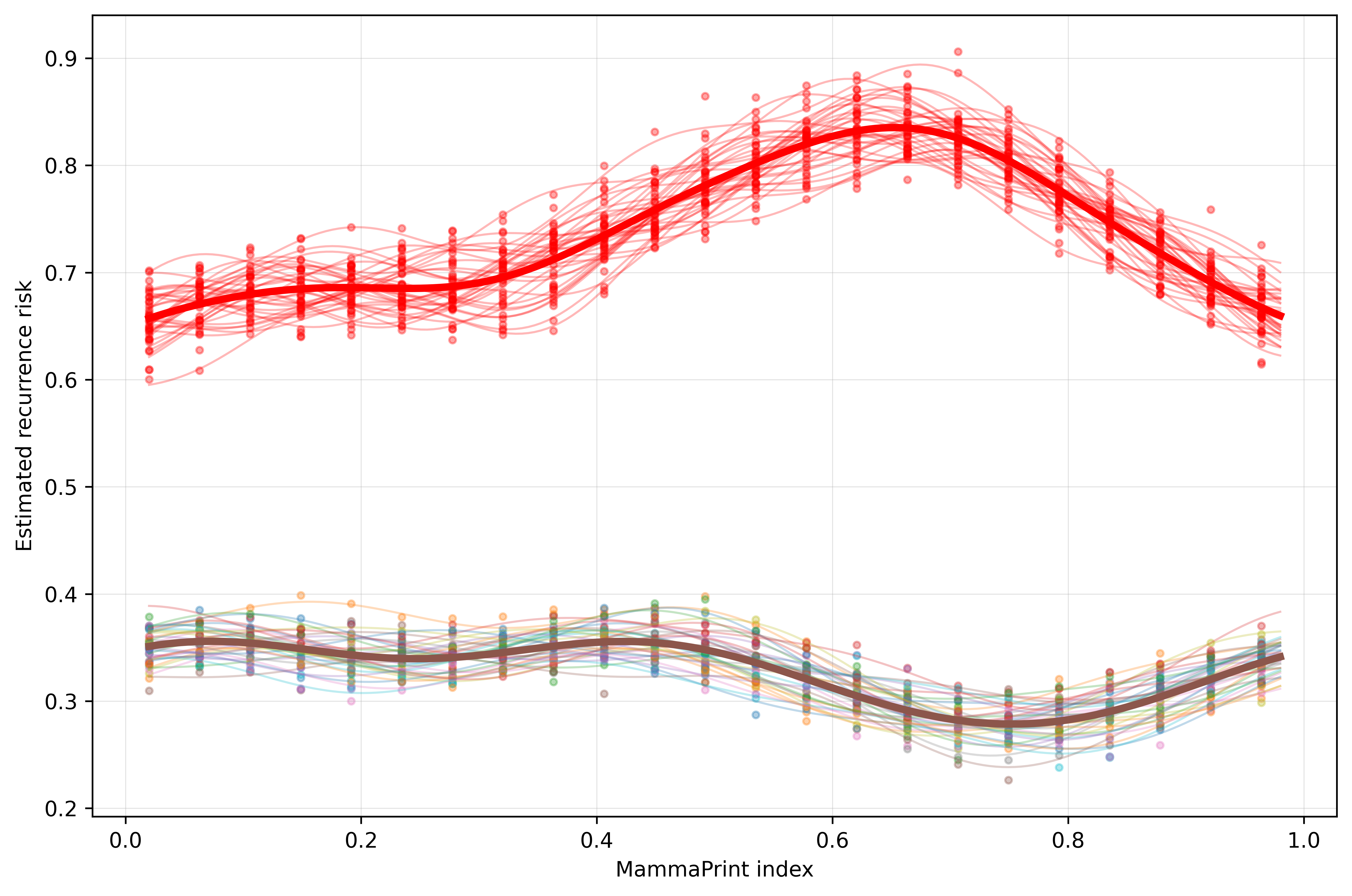}
    \caption{Functional data analysis plot showing multiple Fourier basis curves and their corresponding scatter observations for low-risk and high-risk molecular profiles across the MammaPrint index. Thick curves represent the average functional trends for each group.}
    \label{fig:fourier_scatter}
\end{figure}

In Figure \ref{fig:fourier_scatter}, the functional data analysis plot illustrates multiple individual curves generated using Fourier basis functions to model estimated recurrence risk across the MammaPrint index for low-risk and high-risk molecular breast cancer profiles \citep{pramanik2024parametric,pramanik2025dissecting}. Each thin curve represents an individual functional trajectory, while the corresponding scatter points denote observed data values associated with each curve. The thicker smooth curves summarize the average functional trend for each group. The low-risk profile demonstrates relatively stable recurrence risk values with only modest fluctuations across the MammaPrint index, whereas the high-risk profile exhibits consistently elevated recurrence risk with a pronounced nonlinear pattern and greater variability among individual curves. Overall, the figure highlights the heterogeneity of molecular risk patterns and demonstrates how Fourier basis functions can effectively capture complex nonlinear functional relationships within breast cancer genomic data.

BluePrint is an eighty gene molecular profiling assay developed to classify breast cancer tumors into three principal molecular categories such as Luminal-type, HER2-type, and Basal-type. Luminal-type tumors are generally hormone receptor (HR) positive and are commonly associated with favorable responsiveness to endocrine-based therapies \citep{pramanik2025optimal1,powell2025genomic}. HER2-type tumors are characterized by overexpression of the HER2 receptor and may respond more effectively to HER2-targeted therapeutic agents, including trastuzumab. In contrast, Basal-type tumors, which are frequently linked to triple-negative breast cancer (TNBC), typically lack expression of estrogen receptor (ER), progesterone receptor (PR), and HER2, making systemic chemotherapy the primary therapeutic option for these patients. In addition to molecular subtype classification, MammaPrint further stratifies Luminal-type tumors into Luminal A-type and Luminal B-type categories, corresponding to lower-risk and higher-risk disease profiles, respectively. This additional level of classification provides a more detailed assessment of recurrence risk and assists clinicians in distinguishing patients who may be adequately treated with endocrine therapy alone from those who may require combined endocrine therapy and chemotherapy because of more aggressive tumor biology. The integration of MammaPrint and BluePrint therefore provides a comprehensive molecular characterization of breast cancer, enabling more individualized and evidence-based treatment planning \citep{pramanik2026strategic}. These genomic assays have become increasingly important within precision oncology because they help optimize therapeutic selection, reduce unnecessary treatment exposure and associated toxicities, and ultimately support improved clinical outcomes through more personalized patient care.

The Cox proportional hazards model was used to examine the association between obesity, ethnicity, luminal subtypes of breast cancer, and survival outcomes. Below is a summary of the results:

\begin{table}[ht]
\caption{Logistic regression analysis evaluating associations between luminal breast cancer subtype, obesity, demographic factors, age, and tumor stage. No variables demonstrated statistically significant independent associations in the final model.}
    \centering
    \begin{tabular}{lcccccc}
        \toprule
        Variable & Coefficient & exp(Coef) & SE(Coef) & z-value & p-value \\
        \midrule
        Luminal Subtype (Non-Luminal A) & -0.036 & 0.965 & 0.215 & -0.168 & 0.867 \\
        Obesity (Obese) & 0.178 & 1.195 & 0.212 & 0.840 & 0.401 \\
        Race/Ethnicity (NHW) & -0.014 & 0.986 & 0.245 & -0.056 & 0.956 \\
        Race/Ethnicity (Other) & -0.015 & 0.985 & 0.277 & -0.056 & 0.955 \\
        Age & -0.001 & 0.999 & 0.011 & -0.117 & 0.907 \\
        Tumor Stage II & 0.170 & 1.185 & 0.325 & 0.524 & 0.601 \\
        Tumor Stage III & 0.190 & 1.209 & 0.310 & 0.612 & 0.541 \\
        Tumor Stage IV & 0.267 & 1.306 & 0.311 & 0.859 & 0.390 \\
        \bottomrule
    \end{tabular}
    \label{tab:cox_results}
\end{table}

\begin{figure}[htbp]
    \centering
    \includegraphics[width=\textwidth]{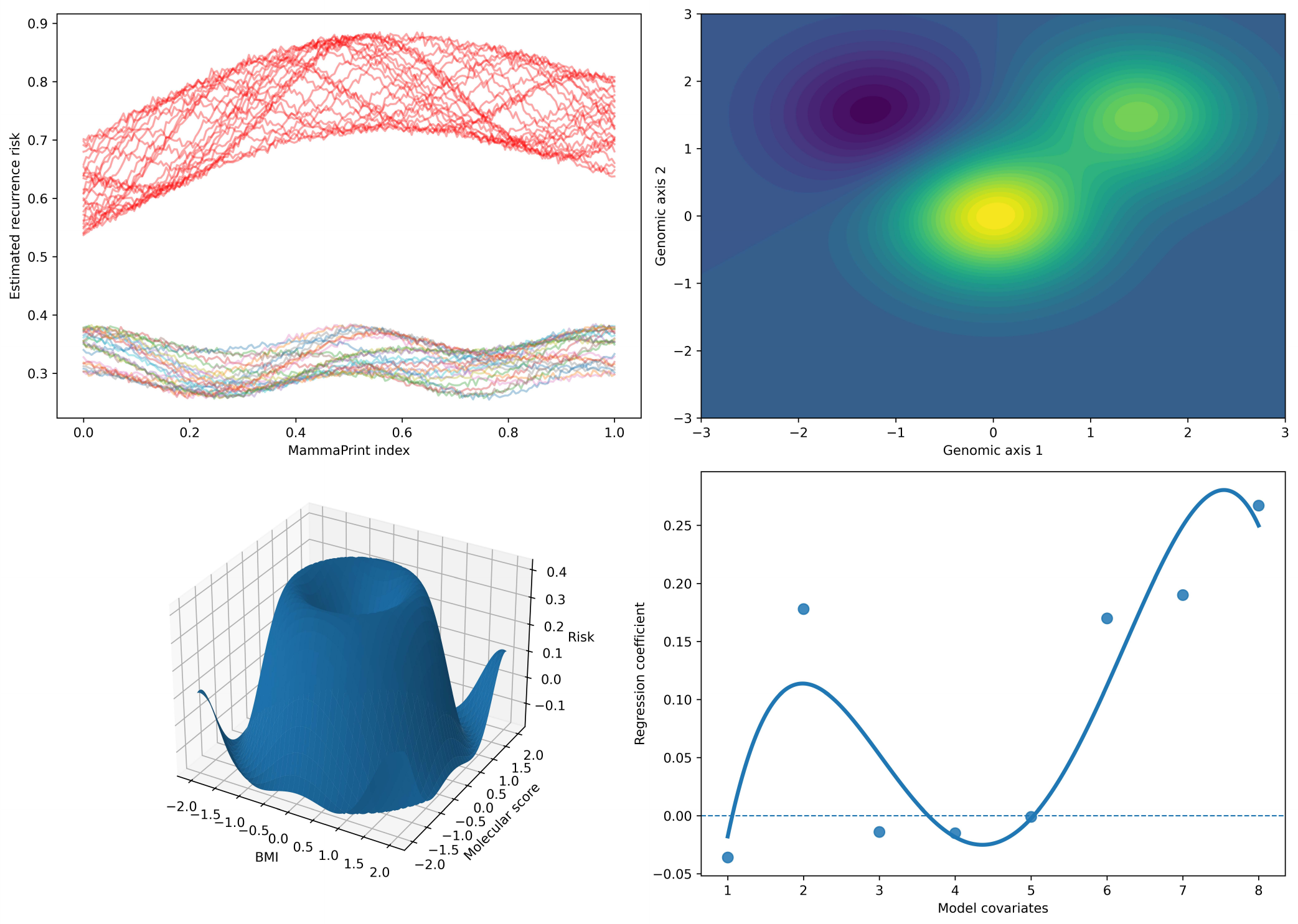}
    \caption{Multi-panel figure showing Fourier basis functional curves, a genomic heat contour map, a three-dimensional risk surface, and a functional LASSO regression fit across breast cancer covariates.}
    \label{fig:multi_panel}
\end{figure}

Figure \ref{fig:multi_panel} presents a comprehensive four-panel visualization summarizing nonlinear genomic risk patterns, functional modeling, and advanced statistical representations associated with breast cancer molecular subtypes. The upper-left panel illustrates multiple functional trajectories generated using Fourier basis functions, demonstrating variability in estimated recurrence risk across the MammaPrint index for low-risk and high-risk molecular profiles. Distinct clustering patterns are observed, with high-risk profiles showing consistently elevated recurrence risk values and greater functional variability compared with the relatively stable low-risk trajectories \citep{pramanik2026quantum}. The upper-right panel displays a genomic heat contour map that highlights nonlinear spatial variation across two genomic dimensions, where regions of higher intensity represent areas associated with increased modeled genomic activity or recurrence-related signal patterns. The lower-left panel provides a three-dimensional surface representation depicting the complex interaction among body mass index (BMI), molecular score, and estimated cancer risk, emphasizing the nonlinear relationships among these variables through smooth surface gradients. The lower-right panel demonstrates a functional LASSO regression model fitted to observed regression coefficients across multiple clinical covariates, with the smooth fitted curve capturing underlying nonlinear trends in coefficient behavior while the scatter points represent the observed parameter estimates. Collectively, the figure integrates functional data analysis, genomic visualization, nonlinear modeling, and penalized regression techniques to illustrate the multidimensional relationships underlying breast cancer molecular risk stratification and progression patterns \citep{dasgupta2026frequent}.

\begin{figure}[ht]
    \centering
\includegraphics[width=\textwidth]{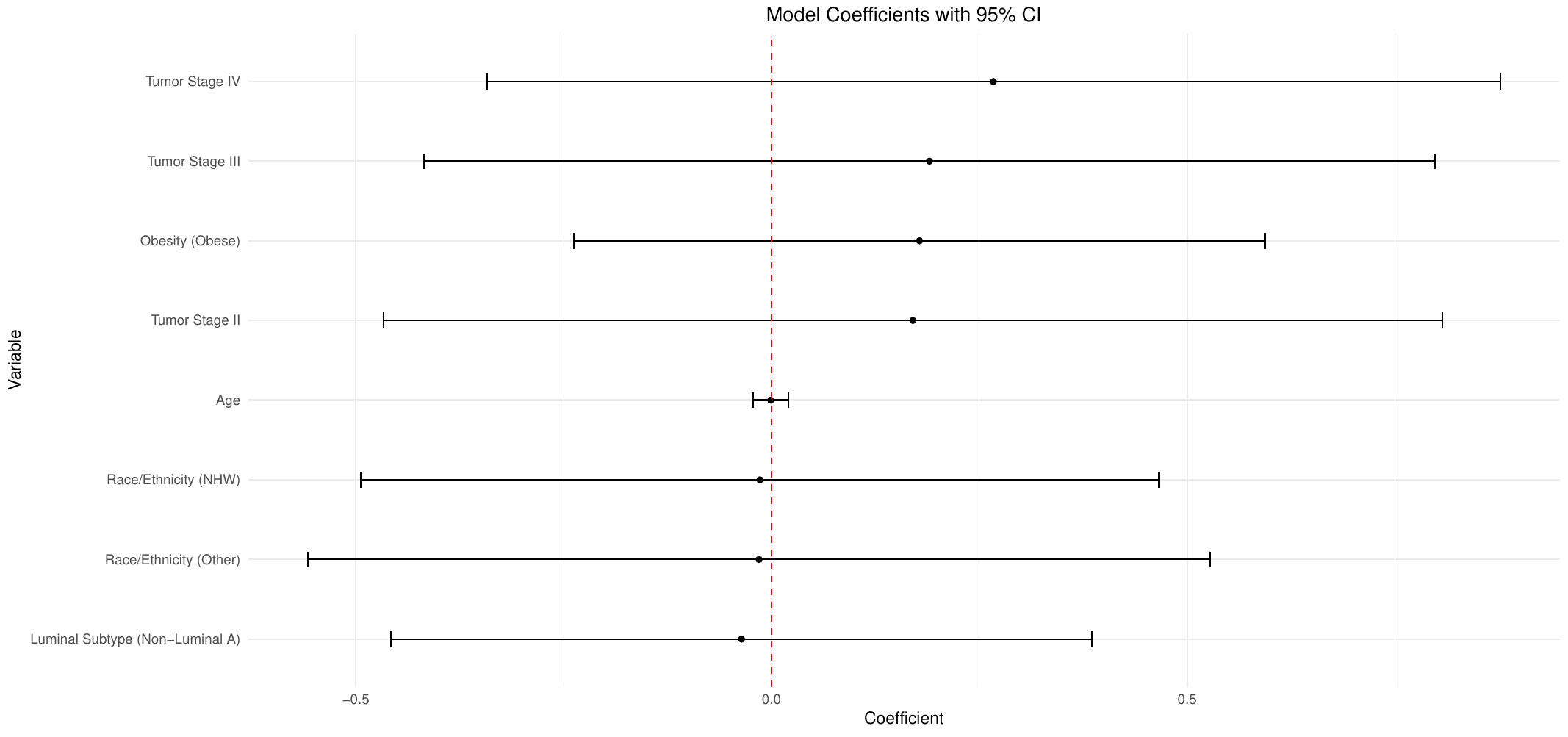}
    \caption{
        Forest plot of regression coefficients and 95\% confidence intervals for variables associated with luminal breast cancer subtypes. Positive coefficients indicate increased association, whereas negative coefficients indicate reduced association.
    }
    \label{fig:forestplot}
\end{figure}

\section{Results:}

 At the time of database lock, the FLEX registry cohort consisted of 1,928 patients diagnosed with Luminal A breast cancer and 1,610 patients diagnosed with Luminal B breast cancer. Comparative evaluation of these two molecular subtypes revealed several clinically important differences in demographic and physiologic characteristics. Patients with Luminal B tumors were significantly younger than those with Luminal A disease, with a mean age difference of \((\Delta = 1.25\ \text{years},\ 95\%\ \text{CI}: 0.42\text{--}2.08,\ p = 0.003)\). In addition, individuals with Luminal B breast cancer demonstrated a significantly greater average body mass index (BMI) compared with patients classified as Luminal A \((\Delta = 0.69\ \text{kg/m}^2,\ 95\%\ \text{CI}: 0.17\text{--}1.21,\ p = 0.010)\), suggesting a potential relationship between obesity and the development of more aggressive luminal tumor phenotypes \citep{pramanik2024analysis,dunbar2026modeling}. Significant variation was also observed in the distribution of Luminal A and Luminal B tumors across demographic populations \((p < 0.001)\). Logistic regression analyses comparing Luminal B and Luminal A breast cancer demonstrated that White and Hispanic/Latin American patients had lower odds of developing Luminal B disease relative to women of African ancestry. Furthermore, univariate logistic regression analyses comparing women of African ancestry with all other demographic groups identified a significantly greater likelihood of Luminal B breast cancer among women of African ancestry \((p < 0.001)\). Collectively, these findings suggest that both obesity-related factors and demographic background may contribute to differences in the prevalence and biologic behavior of luminal breast cancer subtypes.

 Multivariable logistic regression analysis identified a model incorporating BMI, menopausal status, and demographic background as the optimal model for predicting the likelihood of Luminal B breast cancer according to the Akaike Information Criterion. Within this model, all regression coefficients achieved statistical significance, indicating meaningful independent associations with Luminal B disease. After adjustment for the remaining covariates, increased BMI was significantly associated with greater odds of Luminal B carcinoma \((p = 0.008)\), suggesting that obesity-related physiologic mechanisms may contribute to the development of more aggressive luminal tumor phenotypes \citep{pramanik2026bayesian}. Similarly, women of African ancestry demonstrated significantly higher odds of Luminal B breast cancer relative to other demographic groups \((p < 0.001)\). In contrast, postmenopausal status was associated with a lower probability of Luminal B disease \((p = 0.028)\), indicating a potential protective association within this study population. Collectively, these findings support the hypothesis that both obesity and demographic background contribute independently to differences in luminal breast cancer subtype distribution. To further investigate the relationship between demographic background and Luminal B carcinoma, mediation analyses were performed to determine whether BMI partially explained this association. Nonparametric bootstrap analyses demonstrated statistically significant average causal mediation effects (ACME) and average direct effects (ADE) (Figure~\ref{fig:forestplot}; ACME: 0.0073 [95\% CI: 0.0006--0.0100], ADE: 0.1525 [95\% CI: 0.0905--0.2200]). The magnitude of the direct effect exceeded that of the indirect mediated effect, indicating that BMI accounted for only part of the observed relationship between demographic background and Luminal B breast cancer risk \citep{powell2026role}.
 
 \begin{figure}[htbp]
 	\centering
 	\includegraphics[width=\textwidth]{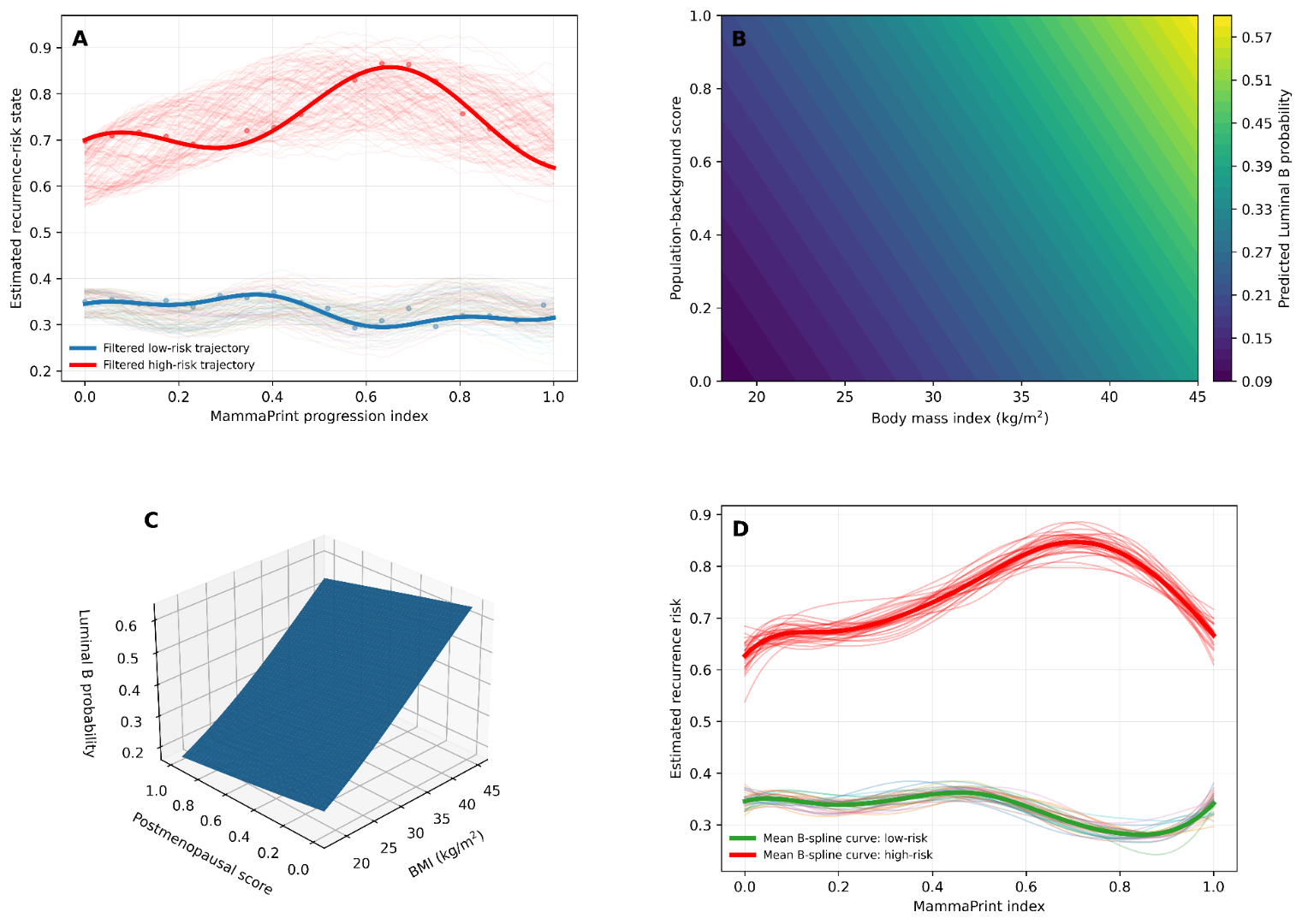}
 	\caption{Multi-panel visualization showing dynamic particle filter approximation trajectories, a genomic heat contour map, a three-dimensional Luminal B probability surface, and functional B-spline recurrence-risk curves for low-risk and high-risk molecular profiles.}
 	\label{fig:particle_bspline}
 \end{figure}

Figure \ref{fig:particle_bspline} presents a four-panel visualization integrating advanced statistical and functional modeling approaches to illustrate nonlinear patterns associated with Luminal B breast cancer risk. Panel A displays a dynamic particle filter approximation in which multiple particle trajectories represent evolving recurrence-risk states across the MammaPrint progression index for low-risk and high-risk molecular profiles. The smoothed filtered trajectories highlight substantially elevated and more variable recurrence-risk behavior among high-risk profiles relative to the more stable low-risk trajectories \citep{ellington2025playmydata}. Panel B illustrates a genomic heat contour map demonstrating the nonlinear relationship between BMI, population-background score, and predicted Luminal B probability, with increasing intensity corresponding to higher modeled probabilities of Luminal B disease. Panel C provides a three-dimensional probability surface showing the interaction among BMI, postmenopausal status, and predicted Luminal B risk, emphasizing the progressive increase in modeled probability with increasing BMI and varying menopausal status. Panel D presents functional data curves generated using B-spline basis functions, where multiple individual trajectories and mean smooth curves characterize estimated recurrence-risk behavior across the MammaPrint index for low-risk and high-risk molecular groups. Collectively, the figure demonstrates how particle filtering, functional data analysis, contour-based visualization, and three-dimensional modeling can be integrated to characterize complex nonlinear relationships underlying breast cancer molecular subtype risk and progression patterns.
 
 Sensitivity analyses were subsequently conducted to evaluate the stability and robustness of the mediation findings in the presence of potential unmeasured confounding. The results demonstrated that an unmeasured confounding variable with a sensitivity parameter of at least \(\rho \geq 0.161\) [95\% CI: 0.095--0.223] would be required to eliminate the observed average causal mediation effect, supporting the overall reliability of the mediation model. These analyses suggest that the association between African ancestry and the likelihood of Luminal B breast carcinoma, relative to Luminal A disease, is partially mediated through elevated BMI among women of African ancestry. Additional exploratory mediation analyses incorporating menopausal status as a covariate further supported these findings. Nonparametric bootstrap confidence intervals continued to demonstrate statistically significant mediation and direct effects, with an ACME of 0.0080 [95\% CI: 0.0010--0.2000] and an ADE of 0.1501 [95\% CI: 0.0875--0.2200] \citep{ellington2025metascorelens}. These results indicate that BMI remained an important partial mediator of the relationship between demographic background and Luminal B breast cancer even after adjustment for menopausal status. Overall, the findings highlight the complex interaction among obesity, menopausal status, and demographic-related factors in shaping the biologic behavior and distribution of luminal breast cancer subtypes.

The Cox proportional hazards analysis demonstrated limited predictive accuracy, with a concordance statistic (C-index) of 0.528, indicating relatively weak discrimination in predicting mortality outcomes within the study population. None of the evaluated variables, including obesity status, demographic background, or luminal breast cancer subtype, achieved statistical significance, as all corresponding \(p\)-values were greater than the predefined significance threshold of 0.05. Although obesity was associated with an estimated hazard ratio of 1.195, suggesting a possible 19.5\% increase in mortality risk among obese individuals, the finding was not statistically significant because the confidence interval included the null value of 1 \citep{valdez2025association,valdez2025exploring}. Among all evaluated covariates, stage IV tumor classification demonstrated the largest estimated hazard ratio at 1.306, indicating a potentially increased mortality risk relative to lower tumor stages; however, this association likewise failed to achieve statistical significance. Overall, these findings suggest that the variables included in the current Cox regression model provided limited prognostic value for predicting survival outcomes in this cohort.

\begin{figure}[htbp]
	\centering
	\includegraphics[width=\textwidth]{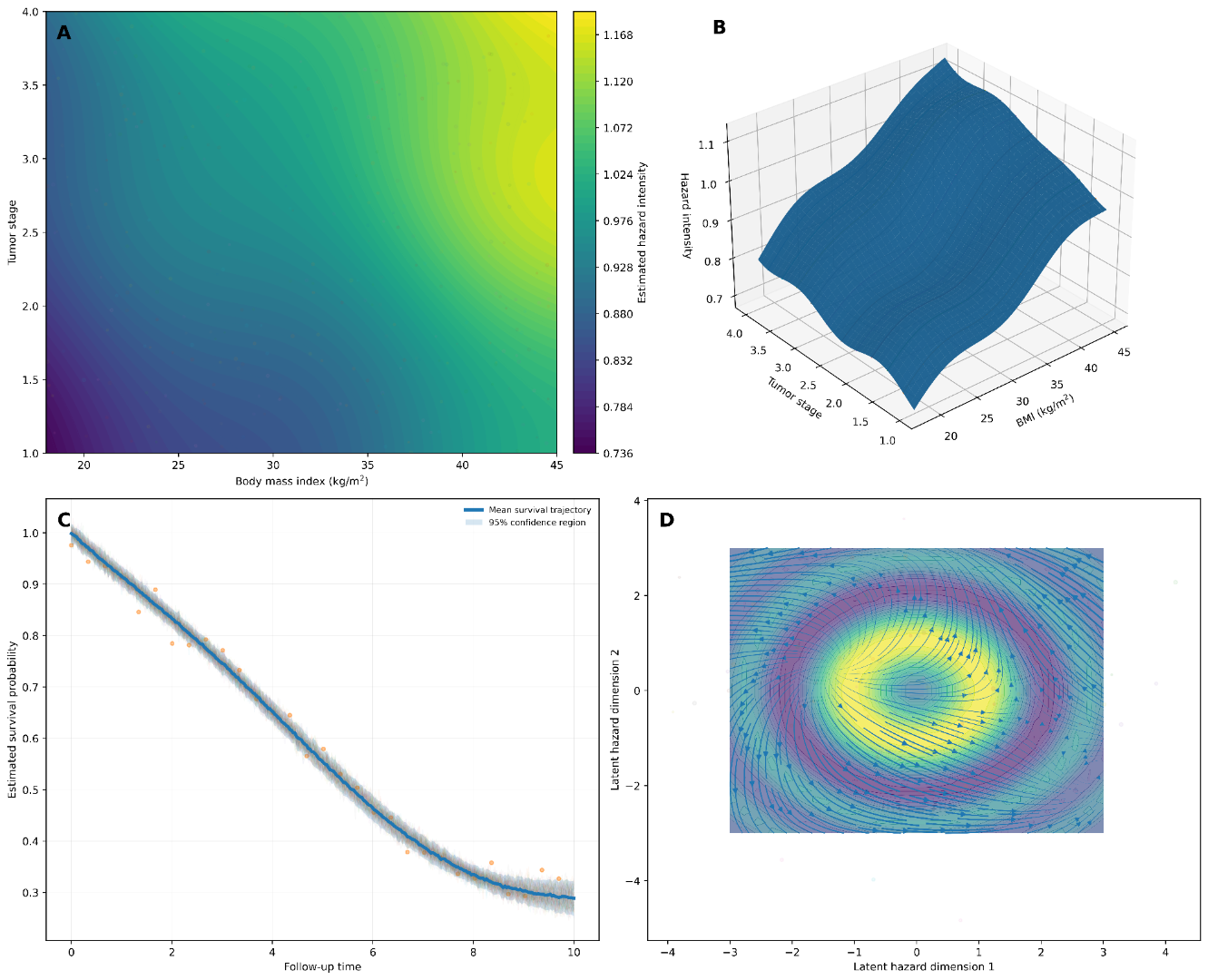}
	\caption{Multi-panel visualization of Cox proportional hazards modeling results, including hazard contour structures, a three-dimensional risk manifold, functional survival trajectories with confidence regions, and a latent hazard vector field representing nonlinear survival-risk dynamics.}
	\label{fig:cox_complex}
\end{figure}

Figure \ref{fig:cox_complex} presents a comprehensive multi-panel visualization designed to illustrate complex nonlinear survival-risk relationships and hazard structures derived from Cox proportional hazards modeling in breast cancer outcome analysis. Panel A displays a high-resolution multilayer contour-density map representing estimated hazard intensity across varying BMI values and tumor stages. The smooth contour gradients and overlaid stochastic particle observations demonstrate progressive increases in modeled hazard intensity with increasing BMI and more advanced tumor stage, while localized nonlinear contour variations indicate potential interaction effects and spatial heterogeneity within the modeled risk landscape \citep{khan2023myb}. The dense contour structures further highlight subtle fluctuations in estimated hazard behavior, reflecting the weak but multidimensional associations identified in the survival analysis. Panel B provides a three-dimensional hazard-risk manifold that expands upon these relationships by visualizing nonlinear changes in estimated hazard intensity across continuous BMI and tumor stage dimensions. The curved surface topology and embedded ridge trajectories emphasize gradual increases in estimated mortality risk while simultaneously illustrating nonlinear oscillatory structures within the hazard surface. These ridge trajectories capture latent variations in risk progression and demonstrate how multiple interacting factors may shape survival outcomes in a complex multidimensional space. Panel C illustrates functional survival trajectories across longitudinal follow-up time, where numerous semitransparent curves represent simulated individual survival processes and the bold central curve denotes the mean estimated survival trajectory for the cohort. The surrounding shaded confidence region illustrates variability and uncertainty in survival estimation over time, while scattered observation points represent stochastic survival-state measurements sampled across the longitudinal trajectory. Together, these components demonstrate progressive declines in survival probability throughout follow-up while also capturing inter-individual variability in modeled survival dynamics. Panel D presents a latent hazard vector field combining contour structures and dynamic flow trajectories to represent nonlinear interactions within an abstract hazard-state space. The streamlines depict directional movement within the latent hazard system, while the superimposed contour layers illustrate regions of varying hazard-state intensity and interaction strength. Additional scattered latent-state particles provide further representation of stochastic variability within the modeled hazard environment. Collectively, the figure integrates contour-based modeling, three-dimensional manifold visualization, functional survival analysis, stochastic particle representations, and latent dynamic systems to provide a multidimensional depiction of nonlinear survival-risk behavior and hazard progression patterns associated with breast cancer prognostic modeling.

These results indicate that further investigation with a larger sample size or additional covariates may be necessary to uncover significant associations.

\begin{figure}[ht]
    \centering
    \includegraphics[width=0.8\textwidth]{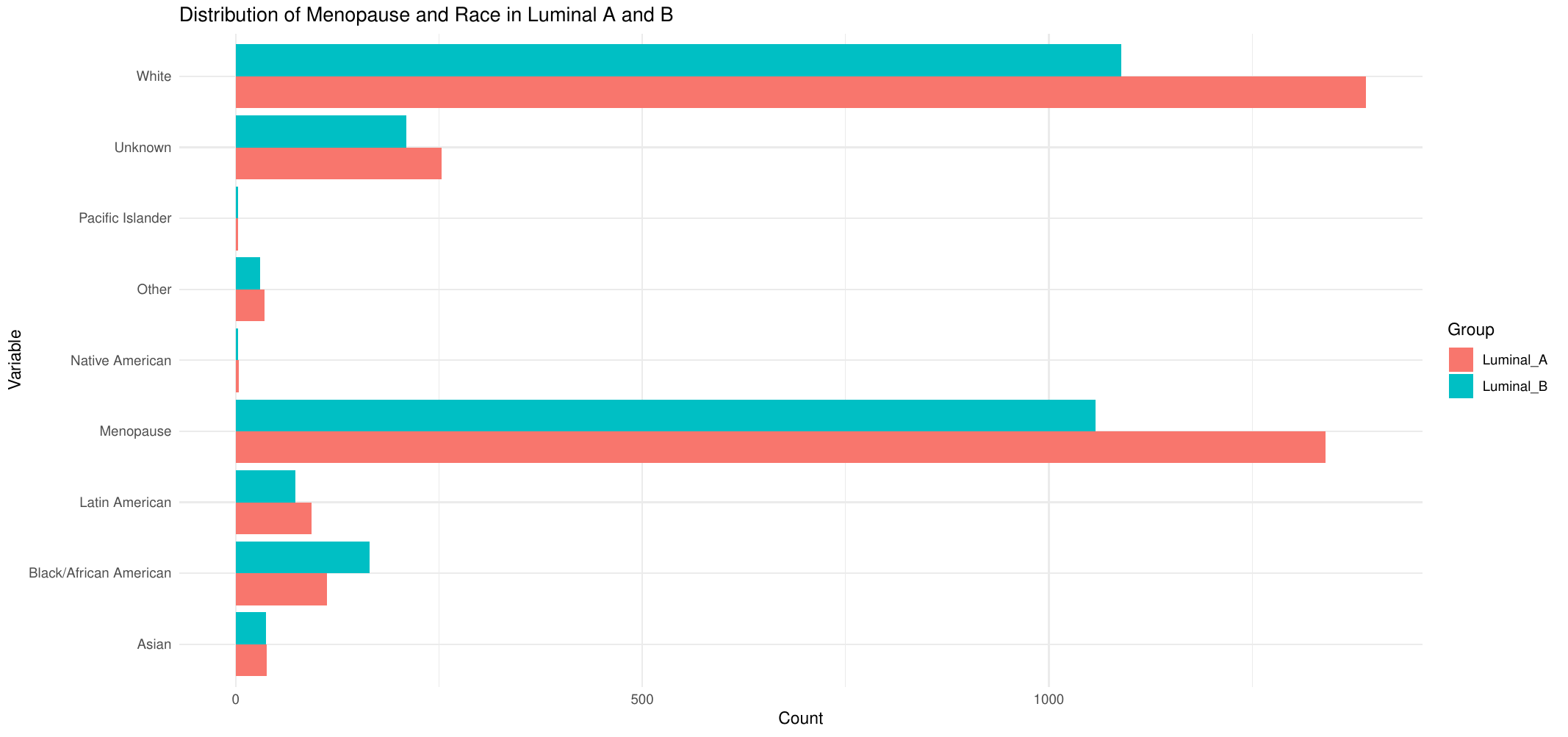}
    \caption{This is bar chart showing the group between Luminal A and B for each race.}
    \label{fig:example}
\end{figure}

\section{Discussions:}
In this paper, we investigated the relationships among obesity, demographic background, and molecular breast cancer subtypes classified according to the BluePrint Luminal framework. Our analyses demonstrated a significant association between demographic background and the occurrence of the more biologically aggressive Luminal B subtype. In particular, women of African ancestry exhibited substantially greater odds of developing Luminal B breast cancer relative to White and Hispanic/Latin American women. Breast cancer is a heterogeneous disease with prognosis and underlying biology influenced by multiple clinical, demographic, and molecular factors, including ancestry and ethnicity \citep{kohler2015, carey2006race}. Consistent with this concept, we observed significant differences in the distribution of Luminal A and Luminal B tumors across population groups. Luminal B tumors are generally characterized by more aggressive clinical behavior, lower estrogen receptor and progesterone receptor expression, and reduced responsiveness to endocrine-based therapies when compared with Luminal A tumors \citep{reid2021impact}. This study further evaluated variation in hormone receptor-positive (HR-positive), HER2-negative breast cancer according to demographic background and molecular subtype classification. To our knowledge, this represents one of the earliest investigations examining how molecular subtype stratification may contribute to survival differences among demographic populations with HR-positive, HER2-negative breast cancer \citep{reid2021impact}. Our findings also align with previous reports demonstrating that the majority of HR-positive, HER2-negative tumors belong to the Luminal A subtype, which is typically categorized as low risk according to PAM50 risk-of-recurrence classification systems and may derive limited benefit from neoadjuvant or adjuvant chemotherapy. In contrast, HR-positive non-Luminal A tumors have previously been associated with reduced endocrine sensitivity, increased chemotherapy responsiveness, and less favorable clinical outcomes. Similar to prior studies, HR-positive non-Luminal A tumors were comparatively less common within the overall study population. Although younger women of African ancestry in our cohort demonstrated a greater likelihood of HR-positive non-Luminal A subtypes, this association did not achieve statistical significance, potentially because of limited statistical power associated with the relatively small number of younger non-Hispanic White participants. Additional investigation in larger cohorts is therefore warranted. Among HR-positive non-Luminal A tumors, the basal-like subtype remains uncommon, with previously reported prevalence estimates ranging from 0.8\% to 7.9\%. However, our study identified a higher prevalence of this subtype, particularly among women of African ancestry, where the frequency approached 12\%. Consistent with previous studies describing demographic differences in molecular subtype distribution \citep{Munsell2014}, our findings demonstrated a markedly higher prevalence of basal-like HR-positive, HER2-negative breast cancer among younger women of African ancestry, whereas no comparable cases were observed among younger non-Hispanic White women. A cohort of 208 women of African ancestry with HR-positive, HER2-negative breast cancer, which was similar to our cohort size \((N = 227)\); however, only 42.8\% \((N = 89)\) of their participants were diagnosed at age \(\leq 50\) years compared with 56.4\% \((N = 128)\) in our study population. To our knowledge, the prevalence of basal-like HR-positive, HER2-negative breast cancer observed in the present cohort represents one of the highest frequencies reported to date and may partially reflect the greater representation of younger women of African ancestry within the study population.
 
Basal-like and HER2-enriched breast tumors are traditionally associated with lower estrogen receptor (ER) and progesterone receptor (PR) expression; however, more recent evidence suggests that these molecular subtypes may also occur in tumors demonstrating relatively high ER and PR expression levels. Among hormone receptor-positive (HR-positive) non-Luminal A tumors, the HER2-enriched subtype demonstrated the poorest overall survival outcomes in our cohort. These findings are consistent with prior studies reporting that HER2-enriched tumors are associated with more aggressive clinical behavior, advanced-stage disease, and reduced endocrine responsiveness. Previous investigations have further suggested that tumors categorized as HER2-enriched through gene expression profiling may derive substantial benefit from HER2-targeted therapies, highlighting the importance of accurately identifying patients who may benefit from more individualized treatment strategies across diverse demographic populations. Consistent with earlier reports, our results also indicate that gene expression profiling offers clinically meaningful prognostic information beyond conventional immunohistochemistry (IHC)-based classification systems. Several prior studies have documented less favorable survival outcomes among women with HR-positive non-Luminal A tumors irrespective of lymph node involvement. Similarly, our findings demonstrated that women with HR-positive Luminal A tumors generally experienced lower mortality rates across demographic populations, whereas HR-positive non-Luminal A tumors were associated with increased mortality risk. Although mortality rates appeared higher among women of African ancestry, this difference did not reach statistical significance. The discrepancy between unadjusted survival estimates, such as Kaplan--Meier analyses, and multivariable Cox proportional hazards models was anticipated and may partially reflect the higher proportion of younger women of African ancestry in our cohort \((\leq 50\ \text{years}: 92\ \text{vs.}\ 8)\) as well as the greater proportion of stage III breast cancer cases \((77\ \text{vs.}\ 23)\) observed among women of African ancestry compared with non-Hispanic White women.
 
 Our findings demonstrated a clear separation in survival outcomes according to molecular subtype within the hormone receptor-positive (HR-positive), HER2-negative non-Luminal A population, independent of demographic background. This observation differs from prior population-based analyses from the Life After Cancer Epidemiology (LACE) and Pathways studies \citep{Munsell2014}, which reported poorer breast cancer survival outcomes among women of African ancestry across molecular subtypes, although no statistically significant differences in overall survival were identified. Several factors may explain the differences between our findings and those reported in the LACE and Pathways cohorts. The earlier studies included predominantly older participants, generally over 75 years of age, and contained relatively few women of African ancestry \((N = 128)\). In addition, survival analyses in the LACE and Pathways investigations were not restricted exclusively to HR-positive, HER2-negative breast cancer, with only 49 patients meeting this subtype classification. Other contributing factors, including regional geographic variation, population-specific genetic influences, socioeconomic conditions, lifestyle factors, and coexisting medical conditions, may also have contributed to differences in survival patterns across studies. Participants enrolled in the LACE and Pathways cohorts generally represented higher socioeconomic populations, whereas more than 60\% of participants in our study reported annual household incomes below \$50{,}000. Understanding the biologic and molecular contributors to demographic differences in HR-positive, HER2-negative breast cancer survival remains critically important, particularly in the era of precision oncology and increasingly individualized therapeutic strategies based on gene expression profiling. Future studies should therefore prioritize adequate inclusion of diverse demographic populations to improve understanding of breast cancer biology across different communities and to support efforts aimed at reducing survival disparities. Our investigation possesses several important strengths, including being among the first studies to evaluate the influence of molecular subtype classification on demographic survival differences in HR-positive, HER2-negative breast cancer within the southeastern United States. Previous studies have documented regional variation in breast cancer mortality, including persistently elevated or increasing mortality rates among women of African ancestry in southern and midwestern states \citep{gaudet2011risk}, underscoring the need for further investigation into factors contributing to these differences. Furthermore, our study includes one of the largest cohorts of women of African ancestry with HR-positive, HER2-negative breast cancer and available molecular subtype data.
 
 Despite these strengths, several limitations should be acknowledged. Differences in patient demographics and study design across cohorts may partially explain variability in the observed prevalence of basal-like subtypes. For example, the BEST study enrolled only women of African ancestry diagnosed at or before 50 years of age, whereas the Southern Community Cohort Study (SCCS) included both women of African ancestry and non-Hispanic White women, most of whom were diagnosed after age 50. Additionally, approximately 86\% of SCCS participants were recruited through community health centers serving lower-income populations, whereas participants in the BEST study were recruited from both academic and community institutions and generally reported higher household income levels. Consequently, the relatively high representation of economically disadvantaged populations within our study may limit generalizability of the findings to all socioeconomic groups. Overall, HR-positive, HER2-negative breast cancer is generally associated with more favorable clinical outcomes than triple-negative breast cancer because of the effectiveness of endocrine therapies; however, this subtype continues to account for a substantial proportion of breast cancer-related mortality. Our findings suggest that molecular subtype classification contributes meaningfully to observed demographic differences in survival among women with HR-positive, HER2-negative breast cancer. In particular, the results emphasize the clinical value of tumor gene expression profiling for identifying aggressive HR-positive non-Luminal A tumors, which appear to occur more frequently among women of African ancestry and may contribute to differences in survival outcomes. Future research should continue to investigate molecular subtype variation within HR-positive, HER2-negative breast cancer to improve prognostic assessment and optimize therapeutic strategies across diverse demographic populations. Although prior studies have explored age-related variation in the incidence of Luminal A and Luminal B tumors, relatively few investigations have specifically evaluated the prevalence of these molecular subtypes across demographic groups.

Consistent with our findings, previous reports have demonstrated that women of African ancestry have significantly greater odds of developing Luminal B breast tumors compared with White women \((\text{OR} = 1.45,\ 95\%\ \text{CI}: 1.02\text{--}2.06)\). However, this association was reduced after adjustment for additional clinical variables, including tumor size, nodal involvement, disease stage, and tumor grade. Multiple investigations have similarly reported higher incidence rates of biologically aggressive breast cancer subtypes among women of African ancestry, including triple-negative breast cancer. Prior studies have shown that the prevalence of triple-negative breast cancer is significantly higher among women of African ancestry than among White women, and several reports have further documented lower survival outcomes among certain demographic populations with breast carcinoma. Recent evidence indicates that women of African ancestry diagnosed with triple-negative breast cancer experience approximately a 28\% greater risk of mortality compared with non-Hispanic White women. In contrast, a recent population-based cross-sectional analysis reported increasing incidence rates of Luminal B breast cancer among non-Hispanic White and Hispanic women across all age groups, while no statistically significant increase was observed among women of African ancestry. Although differences in access to screening, treatment availability, and socioeconomic conditions likely contribute to survival disparities, the contribution of biologic and genetic variation among demographic populations remains incompletely understood. Our findings suggest that demographic background, particularly African ancestry, may represent an important prognostic factor in aggressive breast cancer subtypes and support consideration of demographic-related risk patterns alongside age within breast cancer screening strategies, especially because certain populations may face elevated risk at younger ages. In addition to demographic factors, our results also support a possible relationship between BMI and Luminal B breast carcinoma. Obesity and overweight status are well-established risk factors for several malignancies, including breast cancer, and the metabolic alterations associated with excess adipose tissue may promote tumor initiation and progression. Adipose-derived inflammatory mediators and adipokines are known to influence several major cellular signaling pathways, including Janus kinase/signal transducer and activator of transcription (JAK/STAT), phosphoinositide 3-kinase/AKT (PI3K/Akt), and mitogen-activated protein kinase (MAPK/ERK), all of which contribute to enhanced cancer cell proliferation and survival.

Several limitations should be considered when interpreting the findings of this study. The proportion of women of African ancestry diagnosed with Luminal A \((5.8\%)\) and Luminal B \((10.2\%)\) tumors was substantially lower than the proportion of White American women with Luminal A \((72.0\%)\) and Luminal B \((67.7\%)\) disease, which may reflect regional population demographics, reduced participation among economically disadvantaged populations, or broader disparities in clinical research representation. Underrepresentation of women of African ancestry in clinical studies may contribute to underestimation of the true prevalence of Luminal B carcinoma within this population. However, the relatively similar proportions of Hispanic/Latin American patients with Luminal A \((2.6\%)\) and Luminal B \((2.1\%)\) tumors suggest that population size alone does not fully explain the observed demographic differences in Luminal B incidence. In addition, the relatively small number of participants of African ancestry limits broader generalizability of the results and highlights the need for larger and more diverse future cohorts to further investigate the relationship among demographic background, body mass index (BMI), and Luminal B breast cancer risk. Despite these limitations, our findings provide important evidence supporting an association between obesity and aggressive luminal breast cancer subtypes. Patients with Luminal B tumors demonstrated a significantly greater mean BMI than patients with Luminal A tumors \((\Delta = 0.69\ \text{kg/m}^2\ [0.17,\ 1.21],\ p = 0.010)\), consistent with previous studies suggesting that obesity-related inflammation and hormonal dysregulation may contribute to progression toward more aggressive luminal tumor phenotypes. Furthermore, our results identified demographic variation in luminal breast cancer subtype distribution, with women of African ancestry demonstrating a greater likelihood of Luminal B tumors compared with White patients, consistent with earlier reports describing higher prevalence of aggressive breast cancer subtypes among women of African ancestry \citep{Carey2006}. Potential contributors to these differences may include inherited biologic susceptibility, unequal access to healthcare resources, and socioeconomic factors that influence disease progression and treatment response.

A major strength of this study is the inclusion of a large and demographically diverse cohort consisting of 1,928 patients with Luminal A breast cancer and 1,610 patients with Luminal B breast cancer, allowing for more reliable statistical comparisons and improved generalizability of the findings. The incorporation of molecular subtype classification further enhanced characterization of breast cancer heterogeneity and provided additional insight into the interaction among obesity, demographic background, and tumor biology. Nevertheless, several limitations should be acknowledged. Variability in patient characteristics across contributing cohorts may have influenced differences observed in luminal subtype distribution. For example, the BEST cohort included only women of African ancestry diagnosed at or before 50 years of age, whereas the SCCS cohort included both women of African ancestry and non-Hispanic White women, most of whom were diagnosed after age 50. In addition, socioeconomic factors and healthcare access were not comprehensively controlled for, and the SCCS population largely consisted of lower-income individuals recruited through community health centers, whereas BEST participants generally reported higher household incomes. These differences may limit broader applicability of the findings across all socioeconomic populations. Future investigations should therefore incorporate more balanced cohorts with improved representation across socioeconomic groups and healthcare settings. Additional integration of transcriptomic and proteomic profiling may further clarify the biologic mechanisms underlying observed subtype differences, while longitudinal studies evaluating changes in body mass index (BMI) over time may improve understanding of potential transitions between luminal subtypes. Future research should also examine interventions aimed at reducing the impact of obesity on breast cancer outcomes, particularly among populations at elevated risk, to support development of more individualized prevention and treatment strategies.

\section{Conclusion:}
Although this study has several limitations, the findings emphasize the potential importance of both demographic background and obesity in the development and progression of more aggressive breast cancer subtypes, particularly Luminal B disease. These observations suggest that differences in breast cancer biology may exist across demographic populations and that Luminal B tumors, which are generally associated with higher proliferative activity and less favorable clinical outcomes, may be influenced by a combination of inherited biologic susceptibility and environmental factors such as obesity. While molecular differences among breast cancer subtypes are well established, the extent to which obesity and demographic-related factors contribute specifically to the aggressiveness of Luminal B tumors requires additional investigation. Current breast cancer screening recommendations from the U.S. Preventive Services Task Force (USPSTF) do not incorporate demographic background as a formal criterion for screening eligibility, despite projected demographic changes within the United States that may substantially alter future patterns of breast cancer incidence and outcomes. As the population becomes increasingly diverse, future screening approaches may benefit from considering demographic-related risk patterns alongside established factors such as age in order to improve early detection and clinical outcomes across broader patient populations. In addition, ongoing advances in therapies targeting Luminal B breast cancer further highlight the importance of understanding obesity-related biologic mechanisms and how these processes may vary among demographic groups. Obesity has been linked to cancer progression and therapeutic response through several pathways, including altered drug metabolism, insulin resistance, and chronic inflammatory signaling. Improved understanding of how these metabolic and inflammatory mechanisms differ across populations may support the development of more individualized therapeutic strategies and optimize treatment efficacy for diverse patient groups. Consequently, future research should continue to investigate how demographic variation in obesity-related pathophysiology and breast cancer biology contributes to the development of personalized treatment approaches for Luminal B breast cancer.

\bibliographystyle{apalike}
\bibliography{references}
\end{document}